\documentclass[%
 reprint,
superscriptaddress,
 amsmath,amssymb,
 aps,
floatfix,
]{revtex4-2}

\usepackage{graphicx}
\usepackage{dcolumn}
\usepackage{bm}
\usepackage{xcolor}
\usepackage{hyperref}
\usepackage{multirow}
\usepackage[utf8]{inputenc}
\usepackage[T1]{fontenc}
\usepackage{mathrsfs}
\usepackage{CJK}
\CJKnospace
\usepackage{txfonts}
\usepackage[normalem]{ulem}
\usepackage{tikz}

\newcommand{\brokenlinesegment}{%
\begin{tikzpicture}[scale=0.5]
    \draw[semithick] (0,0) -- (0.2,0);
    \draw[semithick] (0.2,-0.2) -- (0.4,0.2);
    \draw[semithick](0.4,-0.2) -- (0.2,0.2);
    \draw[semithick] (0.4,0) -- (0.6,0);
\end{tikzpicture}}

\begin{document}
\begin{CJK*}{UTF8}{mj}


\title{
Kaleidoscopic reorganization of network communities across different scales
}

\author{Wonhee Jeong (정원희)}
\affiliation{%
    The Research Institute of Natural Science, Gyeongsang National University, Jinju 52828, Korea
}%

\author{Daekyung Lee (이대경)}
\affiliation{%
    Supply Chain Intelligence Institute Austria, Vienna 1030, Austria
}%

\author{Heetae Kim (김희태)}
\email[Contact author: ]{hkim@kentech.ac.kr}
\affiliation{%
    Department of Energy Engineering, Korea Institute of Energy Technology, Naju 58330, Korea
}%

\author{Sang Hoon Lee (이상훈)}
\email[Contact author: ]{lshlj82@gnu.ac.kr}
\affiliation{%
    The Research Institute of Natural Science, Gyeongsang National University, Jinju 52828, Korea
}%
\affiliation{%
    Department of Physics, Gyeongsang National University, Jinju 52828, Korea
}%
\affiliation{%
    Future Convergence Technology Research Institute, Gyeongsang National University, Jinju 52849, Korea
}%

\date{\today}

\begin{abstract}
The notion of structural heterogeneity is pervasive in real networks, and their community organization is no exception. Still, a vast majority of community detection methods assume neatly hierarchically organized communities of a characteristic scale for a given hierarchical level. In this work, we demonstrate that the reality of scale-dependent community reorganization is convoluted with simultaneous processes of community splitting and merging, challenging the conventional understanding of community-scale adjustment. We provide a mathematical argument concerning the modularity function, the results from real-network analysis, and a simple network model for a comprehensive understanding of the nontrivial community reorganization process. The reorganization is characterized by a local drop in the number of communities as the resolution parameter varies. This study suggests a need for a paradigm shift in the study of network communities, which emphasizes the importance of considering scale-dependent reorganization to better understand the genuine structural organization of networks.
\end{abstract}


\maketitle

\section{Introduction}
\label{sec:intro}

Understanding the community structure of networks~\cite{Porter2009,Fortunato_review,Fortunato2022} is a cornerstone of network science~\cite{Barabasi2016,Newman_book,MenczerFortunatoDavis2020}, with implications that span from uncovering hidden relationships within social networks to revealing functional modules in biological systems. The study of network communities offers a window into the mesoscale organization of complex systems, providing insights that are crucial for predicting system behaviors and dynamics. Traditional approaches to community detection have largely been guided by the notion of modularity~\cite{PRE_Girvan-Newman}, a mathematical framework that seeks to maximize the density of links within communities compared to the links between them. However, real-world networks cannot simply be hierarchically categorized, exhibiting dynamic reorganization across different scales. We reveal the intricate process of community reorganization, challenging conventional views, and highlighting the nuanced interplay of community splitting and merging as observational scales change. By investigating the modularity function and analyzing real network data, we illuminate the scale-dependent restructuring that characterizes many complex systems, urging a reevaluation of traditional community detection paradigms.

\section{Modularity Function}
\label{sec:modularity}

A modularity-maximization (MM) method simplifies the complex task of detecting communities into a concrete mathematical objective: finding the partition that maximizes the modularity function~\cite{PRE_Girvan-Newman}
\begin{equation}
Q(\mathcal{G};\gamma) = \frac{1}{M} \sum_g \left( L_g - \gamma \frac{K_g^2}{4M} \right) \,,
\label{eq:modularity}
\end{equation}
where $L_g$ is the sum of weights on the internal edges in group~$g \in \mathcal{G}$, $K_g$ is the sum of the weights on all of the edges connected to the nodes in group~$g$, 
and $M$ is the sum of weights on all edges in the network that plays the role of normalization factor for matching the scales of $L_g$ and $K_g^2$ terms. The resolution parameter $\gamma$ controls the overall scale of communities by weighting the relative importance of internal cohesion of communities~\cite{Reichardt2004}.

This simplification, however, brings at least twofold limitations. First, despite the seemingly benign statement, the maximization process itself is far from being computationally tractable. In fact, a significant portion of development in this field for the past few decades has been dedicated to solving this problem by competitively providing various algorithms to perform the task, where the exact enumeration is next to impossible, even for tiny networks. 
Now, sitting on top of the large pool of such algorithms, we would like to shed light on the second and perhaps larger issue: despite its widespread applications, the modularity function itself definitely has its own limitation in detecting communities~\cite{Peixoto2023} with multiple scales. Even though the aforementioned resolution parameter $\gamma$ is taken as a convenient tool analogous to the adjustment knob of a microscope~\cite{Reichardt2004}, we usually do not know the community scale that we wish to detect \emph{a priori}; thus, this degree of freedom is not always helpful. Moreover, perhaps more profoundly, there is a fundamental limit in detecting communities composed of vastly different scales, i.e., \emph{any} choice of a single $\gamma$ value may not suffice to reveal the most relevant set of communities for a given network. 
In this paper, we focus on a fundamental aspect of this scale-dependence of modularity-maximized communities---what is really happening when $\gamma$ changes? A conventional mantra is a simple progression of hierarchically organized smaller and smaller communities (thus a larger and larger number of communities) as $\gamma$ increases. Our first crucial realization is the fact that the number of communities determines only the average size of communities; it does not provide complete information on the sizes of individual communities that can be genuinely diverse~\cite{JDNoh2005,Lancichinetti2008}, let alone their interwoven structures. 

\begin{figure}
\centering
\includegraphics[angle=0,width=0.8\columnwidth]{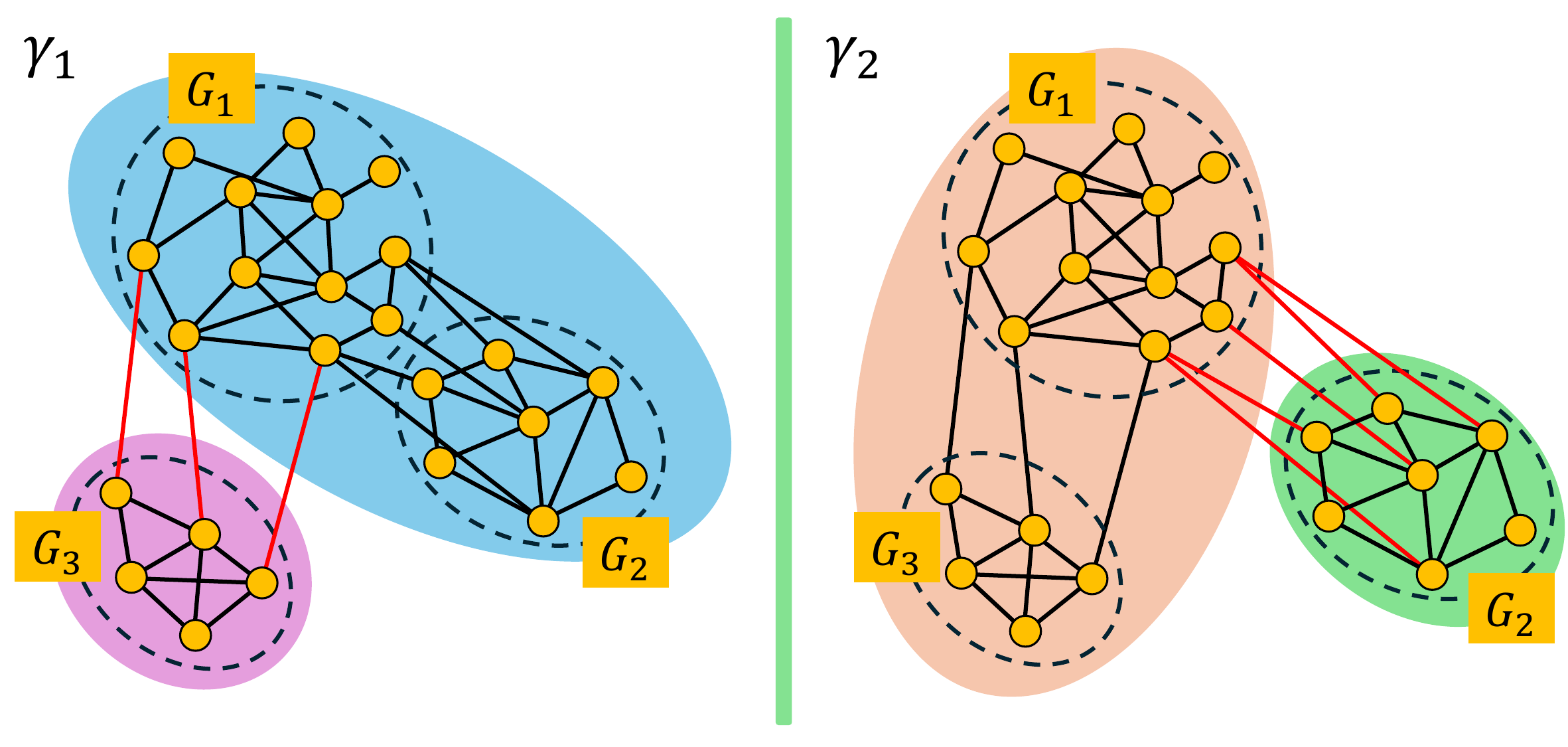}
\caption{A schematic diagram of community reorganization with the fixed number (two) of communities. The left panel has a lower value of $\gamma$ than the right panel, i.e., $\gamma_1 < \gamma_2$. The set of nodes enclosed within the dashed circles indicates that they belong to the same cohesive cluster, and a set with the same background color corresponds to each community for each $\gamma$ value. The edges represented by black and red lines indicate intra- and inter-community connections, respectively.}
\label{fig1}
\end{figure}

\subsection{Change in modularity function at community merging}
\label{sec:Q_change}

Let us take a closer look at the change in modularity for a simple community-merging case. When group $g$ moves into group $h$, one can see that the net change of the modularity in Eq.~\eqref{eq:modularity} is readily calculated by comparing the sum of modularity contributions from separate groups $g$ and $h$ and the modularity contribution from the merged group $g$--$h$,
\begin{equation}
\Delta Q (\mathcal{G} \to \mathcal{G'};\gamma) = Q_{g-h} - \left( Q_g + Q_h \right),
\end{equation}
where $\mathcal{G}' = \mathcal{G}\backslash\{g,h\} \cup \{ g\textrm{--}h \}$.
Suppose that the sum of the weight of edges between nodes in group $g$ and $h$ is $I_{gh}$, and the sum of edge weights connected to outside the $g$--$h$ union from $g$ and $h$ sides are $E_g$ and $E_h$, respectively.
\begin{equation}
Q_{g-h} = \frac{1}{M} \left[ L_g + L_h + I_{gh} - \gamma \frac{K_{g-h}^2}{4M} \right],
\end{equation}
\begin{equation}
Q_g + Q_h = \frac{1}{M} \left[ L_g + L_h - \gamma \frac{K_{g}^2}{4M} - \gamma \frac{K_{h}^2}{4M} \right],
\end{equation}
where $K_g = 2L_g + I_{gh} + E_g$, $K_h = 2L_h + I_{gh} + E_h$, and $K_{g-h} = 2L_g + 2I_{gh} + 2L_h + E_g + E_h = K_g + K_h$.
Thus, the change of the modularity is given by,
\begin{equation}
\Delta Q (\mathcal{G} \to \mathcal{G'};\gamma) = \frac{I_{gh}}{M} - \gamma \frac{K_{g}K_{h}}{2M^{2}},
\label{eq:DelQ}
\end{equation}
where $\gamma$ is the resolution parameter that controls the modularity gain. For the purpose of our argument, consider the community structures for different $\gamma$ values illustrated in Fig.~\ref{fig1}, composed of three cohesive clusters (CCs): $G_1$, $G_2$, and $G_3$ as the basic units (assumed to be not broken up to $\gamma \gg \gamma_2 > \gamma_1$) of community composition. As shown in Fig.~\ref{fig1}, at $\gamma = \gamma_1$, $G_1$ and $G_2$ form a joint community $G_1$--$G_2$, and $G_3$ cannot move into the joint community because 
\begin{equation}
\begin{aligned}
\Delta Q \left( \left\{ G_1,G_2,G_3,\dots \right\} \to \left\{ G_1\textrm{--}G_2,G_3,\cdots \right\} ; \gamma_1 \right) > 0, \\
\Delta Q \left( \left\{ G_1\textrm{--}G_2,G_3,\dots \right\} \to \left\{ G_1\textrm{--}G_2\textrm{--}G_3,\cdots \right\} ; \gamma_1 \right) < 0.
\end{aligned}
\end{equation}
Suppose that at $\gamma = \gamma_2 > \gamma_1$,
\begin{equation}
\Delta Q \left( \left\{ G_1,G_2,\cdots \right\} \to \left\{ G_1\textrm{--}G_2,\cdots \right\} ; \gamma_2 \right) < 0
\end{equation}
and thus $G_1$ and $G_2$ are not merged. In this situation, from the perspective of $G_3$, the question is if it will join the $G_1$--$G_2$ community at $\gamma=\gamma_1$ and if it will join the $G_1$ community at $\gamma = \gamma_2$. From Eq.~\eqref{eq:DelQ}, $I_{(G_1,G_2) G_3} = I_{G_1 G_3}$, $M$, and $K_{G_3}$ are the same, while $K_{G_1-G_2} \to K_{G_1}$ and $\gamma_1 \to \gamma_2$ change; if the inequality
\begin{equation}
\gamma_1 K_{G_1-G_2} > \gamma_2 K_{G_1} \,,
\label{eq:inequality}
\end{equation}
is satisfied, it is possible for both inequalities to hold:
\begin{equation}
\begin{aligned}
\Delta Q \left( \left\{ G_1\textrm{--}G_2,G_3,\dots \right\} \to \left\{ G_1\textrm{--}G_2\textrm{--}G_3,\cdots \right\} ; \gamma_1 \right) < 0 \,, \\
\Delta Q \left( \left\{ G_1,G_2,G_3,\cdots \right\} \to \left\{ G_1\textrm{--}G_3,G_2,\cdots \right\} ; \gamma_2 \right) > 0 \,,
\end{aligned}
\end{equation}
and thus $G_1$ and $G_3$ can be merged at $\gamma = \gamma_2$. 

\begin{figure*}
\centering
\includegraphics[width=\textwidth]{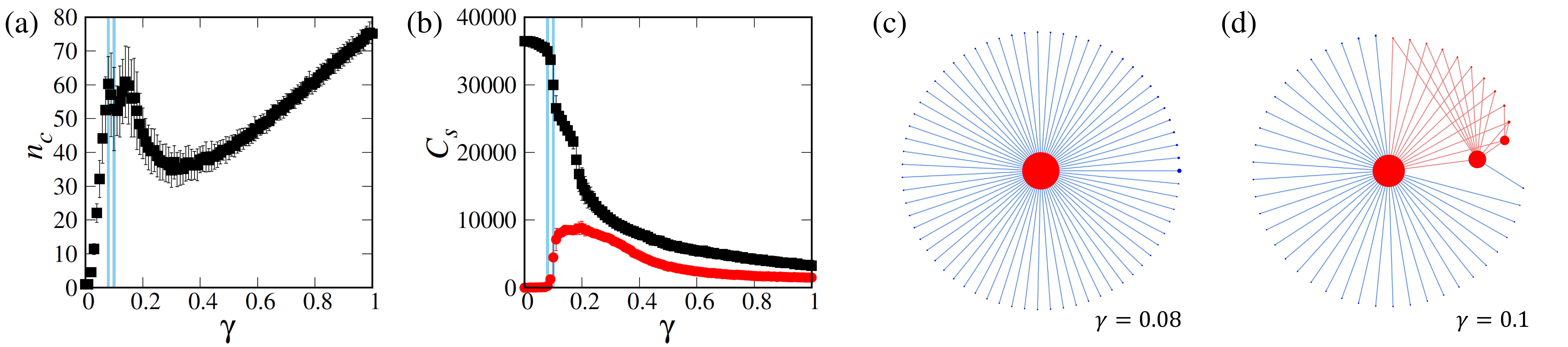}
\caption{(a) The number $n_c$ of communities as a function of the resolution parameter $\gamma$. (b) The sizes of the communities $C_s$ with the first (the black-filled squares) and the second (the red-filled circles) largest sizes as a function of the resolution parameter $\gamma$. For (a) and (b), the results are from the averaged values over $100$ independent community detection results applied to the \texttt{cond-mat} network, and the blue vertical lines indicate $\gamma=0.08$ and $0.1$. The error bars represent the standard deviation, as do all of the other figures in this paper. (c) and (d) show the CGN of communities at $\gamma=0.08$ and $\gamma=0.1$. Each node here indicates a community in the original network, and its size is proportional to the number of nodes composing communities. The edges represent inter-community connections. The communities with red color have two or more neighbors, and the communities with blue color have only one neighbor. The community size is proportional to $(\textrm{the number of nodes in the community} + 5)$ for ideal visibility.}
\label{fig2}
\end{figure*}

\subsection{Community reorganization}
\label{sec:comm_reorganization}

The argument in Sec.~\ref{sec:Q_change} clearly illustrates that the aforementioned conventional viewpoint (larger $\gamma$ values tend to split communities) only considers the separation of $G_1$ and $G_2$ as $\gamma$ grows, while ignoring the possibility of the $G_1$--$G_3$ merging following the $G_1$\brokenlinesegment$G_2$ separation---splitting and merging processes can happen \emph{simultaneously} as $\gamma$ changes and the number of communities is the same as two in both cases of Fig.~\ref{fig1}. In other words, communities in different scales are \emph{reorganized}, not necessarily split/merged unanimously throughout the entire network in response to the $\gamma$ change, as reported in the chromatin interaction network in a biological context~\cite{Bernenko2023}. Suppose more complicated community compositions than Fig.~\ref{fig1}, e.g., a separate hypothetical basic unit or CC $G_4$ at $\gamma = \gamma_1$ that is merged as $(G_1,G_3,G_4)$ at $\gamma = \gamma_2$. Then, the number of communities is \emph{decreased} from three to two for $\gamma_1 \to \gamma_2 > \gamma_1$. 

\section{Results}
\label{sec:results}

\subsection{Real networks}
\label{sec:real_net}

The rather unexpected decreasing phase (DP) or ``dip'' in the number of communities as $\gamma$ is increased, stated in the previous paragraph, is in fact remarkably prominent for $0.15 \lesssim \gamma \lesssim 0.3$ shown in Fig.~\ref{fig2}(a) for the collaboration network between coauthors in the field of condensed matter in Ref.~\cite{cond-mat-2005}, composed of $36\,458$ nodes and $171\,737$ edges weighted by the number of papers. We only take its giant connected component~\cite{Barabasi2016,Newman_book,MenczerFortunatoDavis2020} and use the weighted version of the network for community detection by taking the fully weighted version of adjacency matrix elements and weight sums in Eq.~\eqref{eq:modularity} throughout all of the analysis in this paper. We utilize a representative stochastic MM algorithm called the Louvain algorithm~\cite{Louvain_JSM} and plot the number of communities averaged over $100$ independent realizations for each $\gamma$ value.

A first step to examine the community reorganization process can be the effect of $G_2$ in Eq.~\eqref{eq:inequality}. As long as $G_1$ and $G_2$ are interconnected just enough to form a single community at $\gamma_1$, the relative difference between $K_{G_1-G_2}$ and $K_{G_1}$ would be larger for $G_2$ with a larger size, which in turn would strengthen the degree of inequality. Therefore, even if the number of communities is the same as two for both cases in Fig.~\ref{fig1}, the largest community size tends to decrease by fragmentation of $G_2$ ($G_3$) at a smaller (larger) $\gamma$ value, respectively. The average size of the largest communities against $\gamma$ in Fig.~\ref{fig2}(b) for the \texttt{cond-mat} network strongly indicates such a monotonically decreasing behavior as a function of $\gamma$. Another notable change in the community set is the abrupt growth and slow decay of the average size of the second-largest community. 

Putting things together, we can imagine a pertinent structural change in communities as the following. As $\gamma$ increases from zero, small CCs start to be detached from the largest community (or the sole community at $\gamma \approx 0$), which initiates the increasing number of communities. As the $\gamma$ value increases further, while this detachment process keeps happening, the second-largest community starts to ``absorb'' the detached communities from the largest community, as we will see in detail later. When the rate of the latter exceeds that of the former, the number of communities enters the DP of Fig.~\ref{fig2}(a). In Figs.~\ref{fig2}(c) and \ref{fig2}(d), we compare such a dynamic reorganization of communities near the onset of the DP ($\gamma = 0.08$ versus $0.1$) in terms of their relative size (the number of nodes inside) and connections between them (the edges between the nodes in different communities). 

In particular, the connections between the communities represent a coarse-grained network (CGN) composed of communities and edges connecting them, where two communities are connected if at least an edge exists between their component nodes~\cite{CSong2005}. This CGN of communities helps us to understand the reorganization process; from the single community covering the entire network at $\gamma = 0$, small peripheral communities (not connected to each other) bud out from it when $\gamma$ is slightly increased, which steadily increases the number of communities up to a local maximum at $\gamma = \gamma_*$. This star CGN structure depicted in Fig.~\ref{fig2}(c) starts to deviate by the merged and interwoven peripheral communities as $\gamma$ passes $\gamma_*$ from below [Fig.~\ref{fig2}(d)], which induces the DP until the rate of fragmentation surpasses the rate of merging again. In other words, it is far from a simple uniform fragmentation of communities as $\gamma$ varies, and we claim that this type of highly heterogeneous process prevails in many real networks harboring communities with various scales~\cite{JDNoh2005,Lancichinetti2008}; in many cases, the mixture of communities with multiscale core and periphery separation~\cite{Rombach2017,Kojaku_NJP_cp,WJeong2024} would augment this phenomenon further.

\begin{figure}
\centering
\includegraphics[width=0.95\columnwidth]{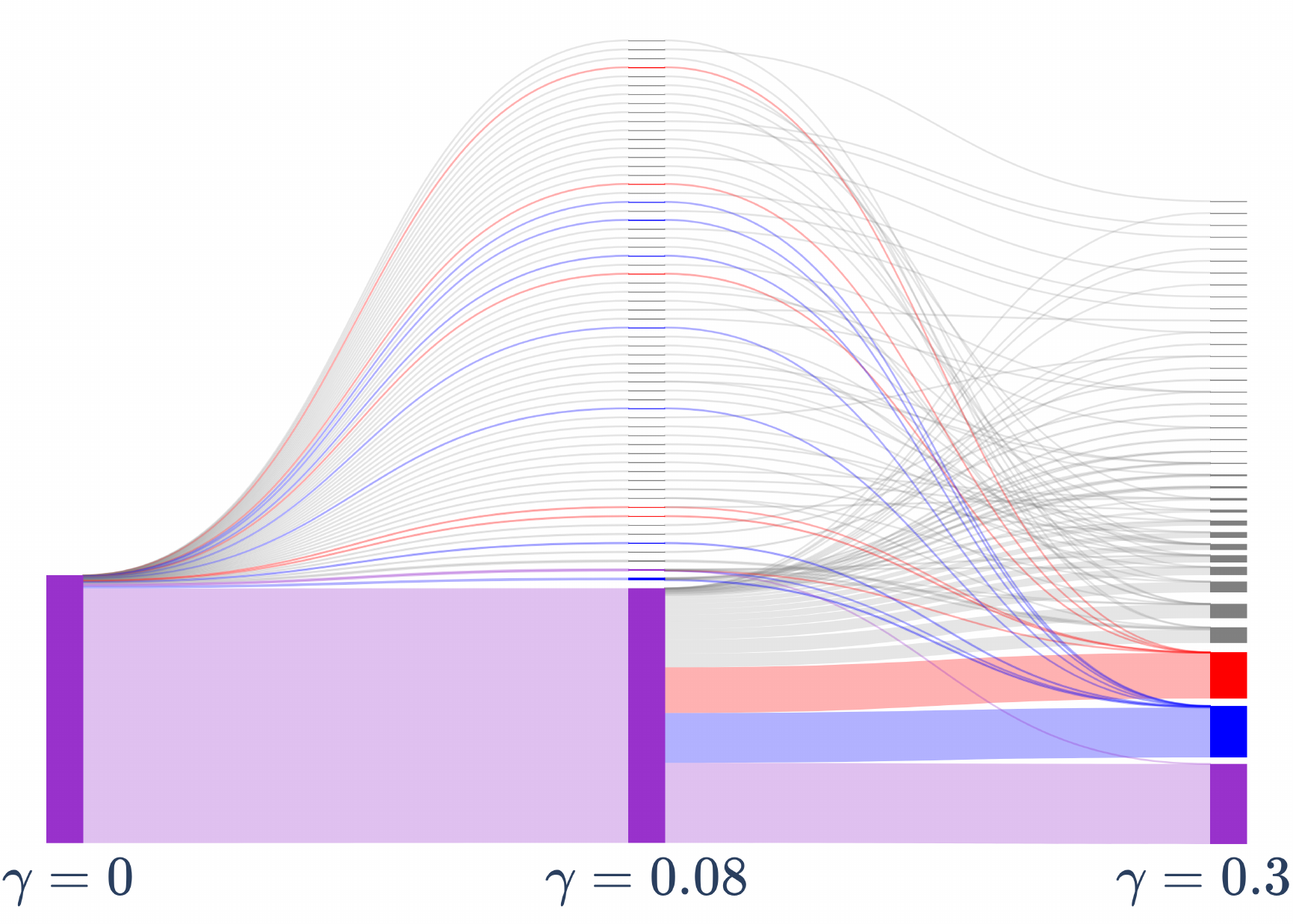}
\caption{The Sankey diagram describes the membership change across communities for different $\gamma$ values. This diagram visually represents the community membership of nodes in a network for given resolution parameters $\gamma$. The flows illustrate nodes' community membership changes as the resolution parameter value varies. The communities are sorted descendingly by the number of nodes from bottom to top, and the three largest communities and their members (and the flows from them) at $\gamma = 0.3$ are colored purple, blue, and red, respectively. The nodes are sorted descendingly by the size of communities to which they belong (from bottom to top) at the one-step larger $\gamma$ value. 
}
\label{fig3}
\end{figure}

So far, we have observed the nontrivial community reorganization by macroscopic observables: the number of communities, the largest and second-largest community sizes, and the coarse-grained networks composed of communities as nodes, from each panel of Fig.~\ref{fig2}. Now, let us take a deeper look at its microscopic aspect represented by nodes' microscopic community membership change over the reorganization process illustrated in Fig.~\ref{fig3}, again in the case of \texttt{cond-mat}. All of the nodes forming the single community covering the entire network for $\gamma = 0$ are roughly separated into the largest community and the rest of the tiny communities surrounding it for $\gamma = 0.08$. The simultaneous process of separating and merging of communities that we mentioned before is well-captured in the flow between $\gamma = 0.08$ and $\gamma = 0.3$ in Fig.~\ref{fig3}; the net change in the number of communities for $\gamma = 0.08$ ($62$ communities) $\to$ $\gamma = 0.3$ ($38$ communities) is negative as the number of small communities or CCs at $\gamma = 0.08$ entering large communities at $\gamma = 0.3$ exceeds the number of segregated communities from the largest communities at $\gamma = 0.08$. 

\begin{figure*}
\centering
\includegraphics[width=\textwidth]{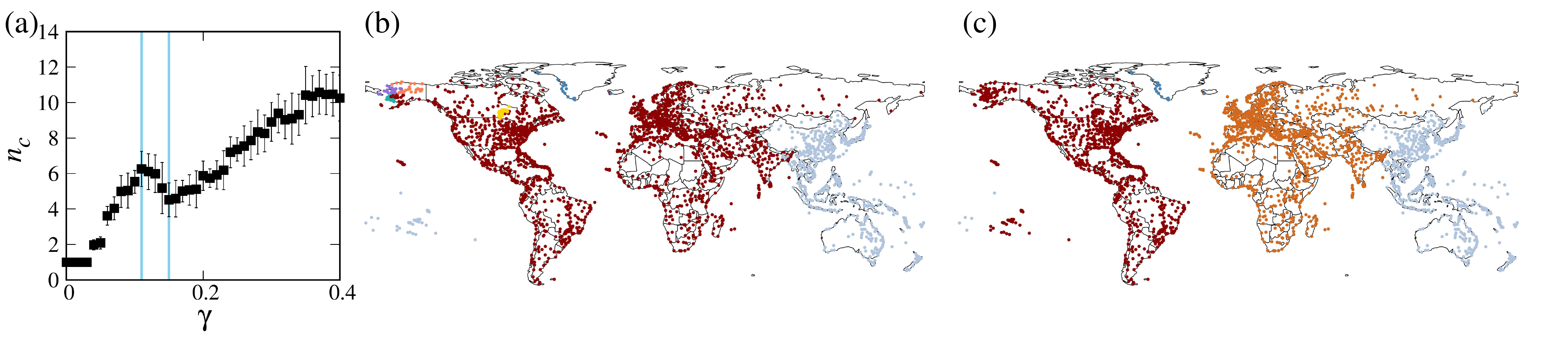}
\caption{(a) The mean number $n_c$ of communities as a function of the resolution parameter $\gamma$, averaged over $100$ community detection results in the case of \texttt{OpenFlights}. We color-code each airport's community membership on the world map, for (b) $\gamma = 0.11$ (seven communities) and (c) $\gamma = 0.15$ (four communities), where the two gamma values are indicated with the blue vertical lines in the panel (a).}
\label{air}
\end{figure*}

We take an airport network (\texttt{OpenFlights}), which is composed of $2\,517$ airports as the nodes and their $18\,068$ interconnections as the edges (weighted by the number of daily flights between two airports)~\cite{OpenFlights}, as another demonstrative example with a clear geographical interpretation.
As depicted in Fig~\ref{air}(a), the number of communities in \texttt{OpenFlights} also shows non-monotonic behavior as $\gamma$ varies, although the decrement is less severe than \texttt{cond-mat}. At $\gamma=0.11$ [Fig.~\ref{air}(b)], AMER (North, Central, and South America) and EMEIA (Europe, the Middle East, Africa, and India) form a single large transatlantic community together, while a small number of airports on Alaska and Canada (A\&C) constitute four separate communities. At $\gamma=0.15 > 0.11$ [Fig.~\ref{air}(c)], however, the separation of AMER and EMEIA communities is accompanied by the absorption of the A\&C airports into the AMER community, which effectively decreases the number of communities from seven ($\gamma = 0.11$) to four ($\gamma = 0.15$). For various types of real networks, of course, the origin of the basic unit of community composition or CC, which is essentially a hyperedge representing higher-order interactions~\cite{Battiston2020}, that we described depends on the systems; it is formed by the papers that authors wrote together for \texttt{cond-mat}, by the geographic proximity for \texttt{OpenFlights}, and by biological units called domains in Ref.~\cite{Bernenko2023}. In any case, though, we believe that similar organizations composed of central large communities and peripheral small communities cause this community reorganization.

\begin{figure*}
\centering
\includegraphics[width=\textwidth]{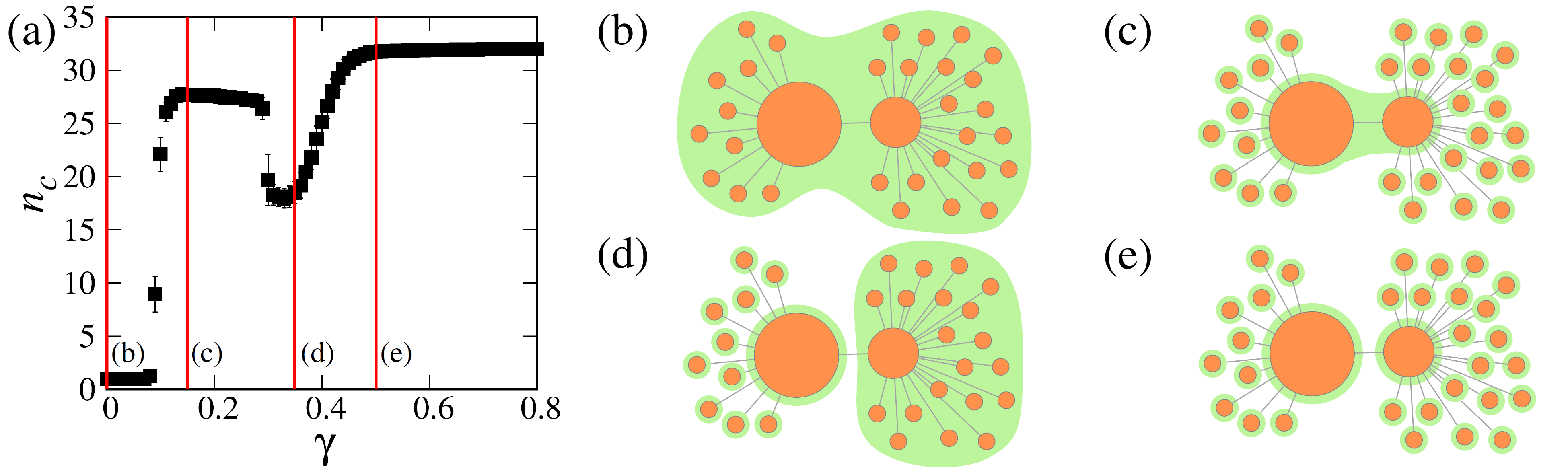}
\caption{(a) The mean number of communities as a function of the resolution parameter in the SBM-based model network, averaging over $100$ independent community detection results from each of $100$ independent network realizations for each $\gamma$ value.
(b)--(e) we show the community division (marked with the green boundary) of the same coarse-grained network composed of elementary building-block communities generated by the SBM at given resolution parameters (b) $\gamma = 0$, (c) $\gamma = 0.16$, (d) $\gamma = 0.36$, and (e) $\gamma = 0.51$, respectively.
}
\label{fig4}
\end{figure*}

\subsection{Stochastic block model}
\label{sec:SBM}

To demonstrate our claim, we suggest a simple generic stochastic-block-model (SBM) type structure~\cite{Karrer2011} to capture the essential part of networks with this behavior. As previously mentioned, there are a few key elements behind it; first, there must be enough heterogeneity in community sizes for non-uniform fragmentation. In addition, the communities tend to be organized as sparsely connected cores and their own peripheries~\cite{Rombach2017,Kojaku_NJP_cp,WJeong2024}, so that the peripheries can be reabsorbed into their cores when their ``master'' cores become smaller by the separation at the core level. Our model network, incorporating those properties, consists of a core part composed of two large communities with $400$ and $200$ nodes, respectively, and $30$ peripheral communities (each community has $20$ nodes); $20$ ($10$) peripheral communities are only connected to the $200$-node ($400$-node) community (no connection between different peripheral communities), respectively. 
The connection probability between the two communities within the core is set to $0.1$, and the connection probability within small and large communities is set to $0.5$ and $0.7$, respectively.
For each peripheral community, $20$ edges (one edge per node in the peripheral communities) exist between it and its master core community. With this restriction, all of the edges are drawn uniformly at random~\cite{Erdos1959}. This composition is precisely described in Figs.~\ref{fig4}(b)--\ref{fig4}(e), where the algorithm finds the exact prescribed communities at the particular $\gamma$ value.

From this SBM, we successfully obtain this precise behavior of community reorganization, as shown in Fig.~\ref{fig4}. The number $n_c$ of communities as a function of $\gamma$ is non-monotonic [Fig.~\ref{fig4}(a)] with local decrement starting from the single community at $\gamma = 0$ [Fig.~\ref{fig4}(b)]. The characteristic pattern of superstar-network formation is observed at $\gamma = 0.16$ [Fig.~\ref{fig4}(c), with a local maximum of $n_c$], and the larger core community ($400$ nodes) absorbs its peripheral communities at $\gamma = 0.36$ [Fig.~\ref{fig4}(d), with a local minimum of $n_c$) as the A\&C airports merged into the AMER community in Fig.~\ref{air}(c). When $\gamma$ increases more, the peripheral communities of the larger core community are split from the core again, as shown at $\gamma = 0.51$ [Fig.~\ref{fig4}(e)]. The result demonstrates that this simple set of ingredients is enough to generate this phenomenon of simultaneous splitting and merging processes, and we suggest that real networks including similar (sub-)structures would yield it. 

\section{Summary and Discussion}
\label{sec:summary}

In summary, we have investigated the nontrivial reorganization of network communities detected by MM, across different scales. The most crucial finding is the simultaneous processes of splitting and merging of communities when the resolution parameter $\gamma$ varies in the modularity function~\eqref{eq:modularity}, while the conventional notion of $\gamma$ tends to be taken just as a practical tool to adjust the overall scale of communities. We have stressed that the simple expression: `scale of communities,' in fact, disguises the complicated reorganization of communities, and our main message in this paper is to explicitly unleash the overlooked reality by both the mathematical argument in principle and the actual observation in practice. Our observation suggests that in real networks composed of large core communities with a multitude of CCs surrounding them, the relative rate of fragmentation and merging can be overturned even when $\gamma$ monotonically varies, which is the key element of the seemingly puzzling local minimum in the number of communities. This mechanism is supported by the size trajectories of the largest and second-largest communities and the CGN. Finally, we have successfully demonstrated that a simple SBM-type network is enough to recapture this phenomenon. 

Beyond this finding and explanation, we are fully aware of the problematic aspects of MM methods~\cite{Peixoto2023}; our aim is more on a practical side---this observed reorganization of communities may be an artifact of MM methods with a single $\gamma$ value at each realization, but we focus on what is really happening in the landscape of community division that this highly popular method generates, and what we can learn from it instead of just dodging the method altogether. Our outlook based on what we have learned throughout this work is the following.
Since real networks are entangled with diverse origins and complex constraints, sufficiently large networks (in terms of the number of nodes) can have numerous CCs that constitute a nontrivial aggregation of communities. In that case, the scale-dependent reorganization of communities is likely observed. The observation of hierarchical structures in previous studies~\cite{hierarchy1,hierarchy2}, rather than reorganization, can be caused by the networks' small size or their structural simplicity. Therefore, we strongly suggest that studies on community structures of (large-scale) networks should consider this scale-dependent reorganization, and there can be other frameworks to study this aspect. As a more practical note, for instance, investigating more detailed mechanistic relationships between CCs and reorganization will be an important challenge in this field.

\begin{acknowledgments}
The authors appreciate Ludvig Lizana and Hye Jin Park (박혜진) for the insightful discussions and comments during S.H.L.'s visit to the Integrated Science Lab (IceLab) in Ume{\aa} University. 
This work was supported by the National Research Foundation (NRF) of Korea under Grant No.~NRF-2021R1C1C1004132 (S.H.L.), No.~NRF-2022R1A4A1030660 (W.J. and S.H.L.), and No.~NRF-2022R1C1C1005856 (D.L. and H.K.). 
\end{acknowledgments}

\end{CJK*}

\bibliography{References}
\end{document}